# Baseline-free scanned-wavelength direct absorption spectroscopy: theoretical analysis and experimental validation

Yihong Wang

*School of Energy and Environment, Southeast University, Nanjing 210096, China*

email wyh@seu.edu.cn

**Abstract** Applicability and accuracy of traditional scanned-wavelength direct absorption spectroscopy (SDAS) diagnostic method is largely affected by combustion state such as temperature and pressure. To resolve this problem, an innovative baseline-free scanned-wavelength direct absorption spectroscopy (BF-SDAS) method is proposed in this work. This method does not need the zero-absorption regions for baseline fitting and can exclude the influence of baseline fitting error on the final result. It is especially suitable for the cases when the scanning range is too narrow or the absorption linewidth is too large to acquire the baseline. In this paper, the BF-SDAS is first derived and theoretically analyzed based on the Fourier analysis, and then the accuracy of this method is validated by measuring the concentration of $H_2O$ molecules in a static gas absorption cell at room temperature.

## 0 Introduction

Tunable Diode Laser Absorbance Spectroscopy (TDLAS) technology has broad application prospects in combustion diagnostics, not only due to its capability to instantaneously acquire key combustion parameters, such as concentration, temperature, pressure and velocity, but also its characteristics of high sensitivity, quick time response, and non-intrusive measurement [1-8]. Scanned-wavelength direct absorption spectroscopy (SDAS）is the most widely used category of TDLAS. Traditional SDAS method is quite sensitive to the baseline fitting error. Particularly in high-pressure combustion environment, the absorption line broadens and makes it difficult or even impossible to obtain desired zero-absorption fitting baseline. Therefore, the range of scanning wavelength has certain requirements in order to obtain accurate and complete absorption spectrum. However, modulation range of widely used diode lasers for communication is usually within 2cm$^{-1}$ [6], and it decreases with the increase of modulation frequency. To obtain complete absorption spectrum, the modulation range of diode laser is constrained to be less than 10 kHz. This limits the use of SDAS in high-pressure environment, and the time response rate is generally less than 10 kHz.

Compared with other diagnostic techniques, the SDAS system is easy to implement and data processing method is also simple [5-7, 12, 16], making it a priority choice among all the TDLAS techniques. However, in some cases the difficulty of obtaining correct baseline and consequently the complete absorption spectrum, limits the application of SDAS. Therefore, this paper proposes a new processing method, baseline-free scanned-wavelength direct absorption spectroscopy (BF-SDAS), to further extend the applicability of SDAS and overcome the shortcomings of traditional SDAS to baseline sensitivity. In the following of this work, the BF-SDAS method is first deduced based on Fourier analysis, then the method is validated by measuring the $H_2O$ concentration in a static gas at room temperature, where the 7185.6 cm$^{-1}$ absorption line of $H_2O$ molecule is used.

## 1 Theoretical background

### 1.1 Beer-Lambert law

When a semiconductor laser with irradiance $I_0$ transmit through the gas to be measured, the laser can be absorbed if the laser frequency $v$ is the same as the transition frequency of the gas absorption species. Denote the laser irradiance after transmission as $I_t$, transmittance $\tau_v$ and absorbance $\alpha_v$ can be defined as following [1-5, 7-8, 12],

$$\tau_v = \left(\frac{I_t}{I_0}\right)_v = \exp(-k_v L) = \exp(-PXS(T)\emptyset(v)L) \tag{1}$$

$$\alpha_v = -\ln\left(\frac{I_t}{I_0}\right)_v = k_v L = PXS(T)\emptyset(v)L \qquad (2)$$

where $k_v$ [cm$^{-1}$] is absorption coefficient, L[cm] is the optical path, P [atm] is pressure, X [mol/mol] the relative molar concentration of absorption specie, $\emptyset(v)$ the line-shape function satisfying $\int_{-\infty}^{+\infty}\emptyset(v)dv = 1$, S(T) [cm$^{-2}$/atm] the intensity of the absorption line at temperature T[K].

Variation of absorption intensity with temperature obeys the following relation [9-11],

$$S(T) = S(T_0)\frac{Q(T_0)}{Q(T)}\frac{T_0}{T}\exp\left[-\frac{hcE''}{k}\left(\frac{1}{T}-\frac{1}{T_0}\right)\right]\left[1-\exp\left(\frac{-hcv_0}{kT}\right)\right]\left[1-\exp\left(\frac{-hcv_0}{kT_0}\right)\right]^{-1} \qquad (3)$$

Where h[J.S] is Planck constant, c[cm/s] is speed of light in vacuum, k[J/K] the Boltzmann constant, E'' [cm$^{-1}$] the energy at lower level, $T_0$(=296 K) the reference temperature, $v_0$ the line central frequency, Q(T) the partition function of the absorbing molecule, which can be calculated using method in reference [13]. Integrate the absorbance $\alpha_v$ in frequency domain and use $\int_{-\infty}^{+\infty}\emptyset(v)dv = 1$ relation, we get the integrated absorbance,

$$A = \int_{-\infty}^{+\infty}\alpha_v\,dv = PXS(T)L \qquad (4)$$

**1.2 SDAS method**

The SDAS method modulates the output of laser generator with saw-tooth modulation signal, to allow scanning of the entire absorption shape function. Directly detected signal is the transmitted light intensity signal $I_t(v)$ attenuated by the gas under investigation. It is necessary to eliminate the influence of light intensity and obtain a light intensity independent absorption spectrum signal. Using data fitting, the zero-absorption regions of scanned wave is used to fit the light intensity baseline $I_0(v)$. For a typical saw-tooth type scanned wave, the light intensity baseline fitting is shown in Figure 1. The laser intensity is under a thresholding value for the initial part of the scanned saw-tooth wave, which can be used to remove the environment background radiation and electric current background in the circuits [5, 14]. Using Eqn. (2), the absorbance $\alpha_v$ can be calculated, as shown in Figure 2. If the temperature is known, the molar concentration of the absorbing species can be calculated with equation (5) [5, 8],

$$X = \frac{A}{PS(T)L} \qquad (5)$$

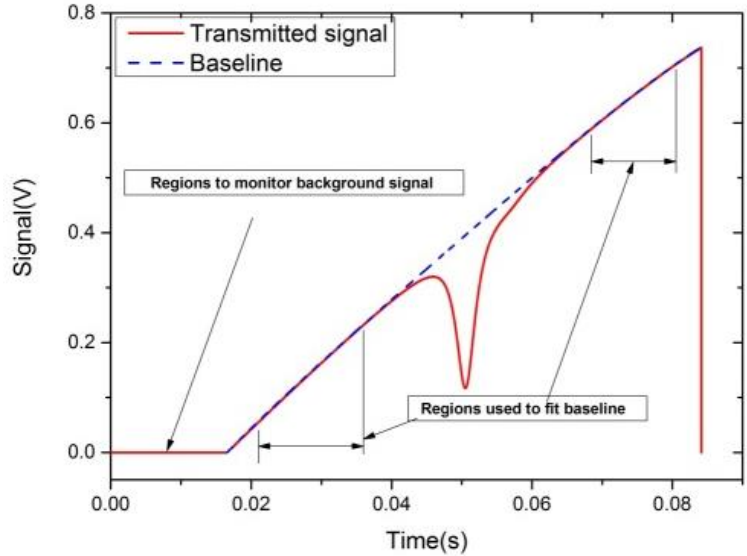

Fig. 1 Illustration of baseline fitting in a typical SDAS

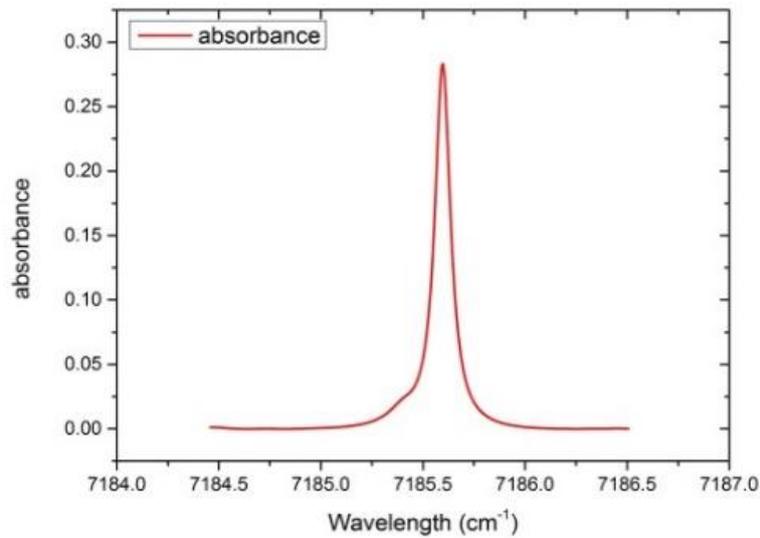

Fig. 2 Absorption rate calculated

## 2 Deficiency of traditional SDAS

Traditional SDAS method is extremely sensitive to errors in baseline fitting. For instance, a 1% baseline fitting error would cause a 10% measurement error, assuming the absorbance is 10%. Figure 3 shows a simulation results of $H_2O$ in the vicinity of 7185.6cm$^{-1}$ spectral line. Since there is no ideal zero-absorption regions, the baseline in panel (a) obtained by polynomial fitting is always smaller than the true initial light intensity, which leads to measured spectral absorbance in panel (b) lower than the true value, because the measured signal is sensitive to the absolute absorption level. For weak absorption application, measured signal-to-noise ratio becomes worse [5, 17-18].

In high pressure combustion, pressure broadening causes overlapping of spectral lines, making the baseline fitting very difficult or even not available [5,6]. Figure 4 shows the simulation results of $H_2O$ in the vicinity of 7185.6cm$^{-1}$ spectral line under different pressure. It can be seen that absorption spectrum varies intensively with pressure rise. For instance, the width of absorption spectrum at P= 1atm is around 1 cm$^{-1}$, while for the case with $P$= 5 atm, the wings of absorption is non-zero, i.e. non-zero absorption wings. In this case, laser modulation range must be enlarged in order to obtain the complete absorption spectrum of this line. However, the modulation range of modern semiconductor laser is usually within 2 cm$^{-1}$, so the complete absorption spectrum of this absorption line cannot be obtained by SDAS technique anymore.

It is necessary to operate the laser at a high modulation frequency [18], in order to ensure suitable transient response characteristics in case of severe variation of gas state (such as in measurements of gas parameters in a combustion environment). However, with the increase of the modulation frequency, the modulation range of the semiconductor laser will be further reduced, which is contradictory to the aim of obtaining complete absorption spectrum, limiting the applicability of SDAS technology.

In short summary, traditional SDAS is only suitable for applications where the absorbance is appropriate (too large or too small will reduce the measurement sensitivity), the pressure is relatively low, the modulation frequency is not large, and the absorption line is relatively isolated.

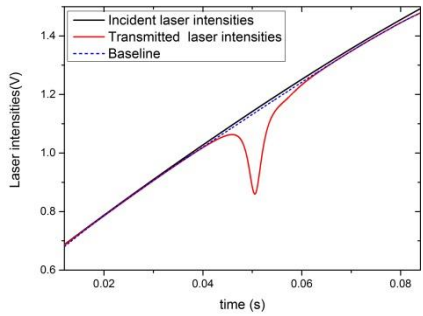 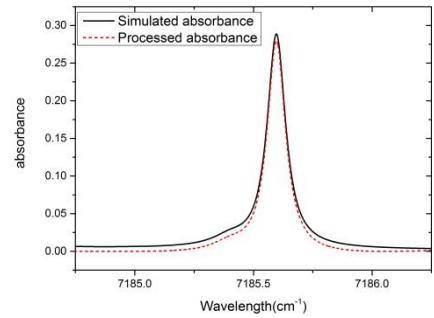

(a) (b)

Fig. 3 (a) simulation of SDAS laser intensity and baseline fitting (b) actual absorbance and simulated absorbance from baseline fitting (simulation condition: 2% $H_2O$; T = 298 K; L = 100 cm; P=1 atm )

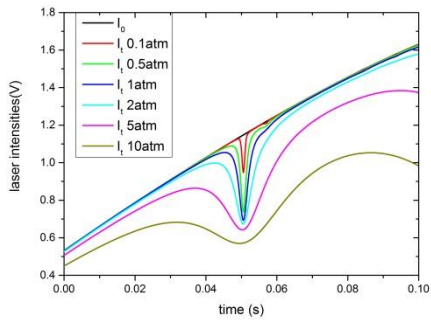 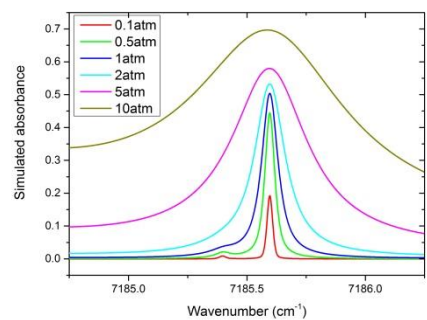

(a) (b)

Fig. 4 (a) transmitted laser intensity under different pressure (b) simulated absorbance variation under different pressure (simulation condition: 2% $H_2O$, T = 373 K, L = 100 cm, P=0.1atm, 0.5atm, 1atm, 5atm, 10atm)

## 3 Fourier based BF-SDAS method

### 3.1 FFT of simulated absorption signal

Figure 5(a) shows the incident and transmitted laser intensity of a typical SDAS, and Figure 5 (b) shows the FFT corresponding amplitude within a certain range. Obvious overlapping of incident and transmitted laser signal can be found on the frequency spectrum of Figure 5 (b), i.e. it is indistinguishable between incident and transmitted signal. To distinguish these two, Nuttall time window is applied to the time domain signals, and the wave form is shown in Figure 5(c). After the Nuttall time windowing operation, the absorption of the middle part of the signal has been strengthened while the edge part is almost zero. Figure 5(d) then shows the FFT amplitude spectrum of the windowed signals. Apparent difference can be observed in the FFT of windowed incident and transmitted signals. For the part between 40~400 Hz in particular, amplitude of transmitted signal is 100 times the incident intensity, which means the baseline fitting error can be reduced 100 times. This simulation results illustrates that Nuttall windowing operation can effectively overcome the traditional SDAS technology to the shortcomings of the baseline sensitivity.

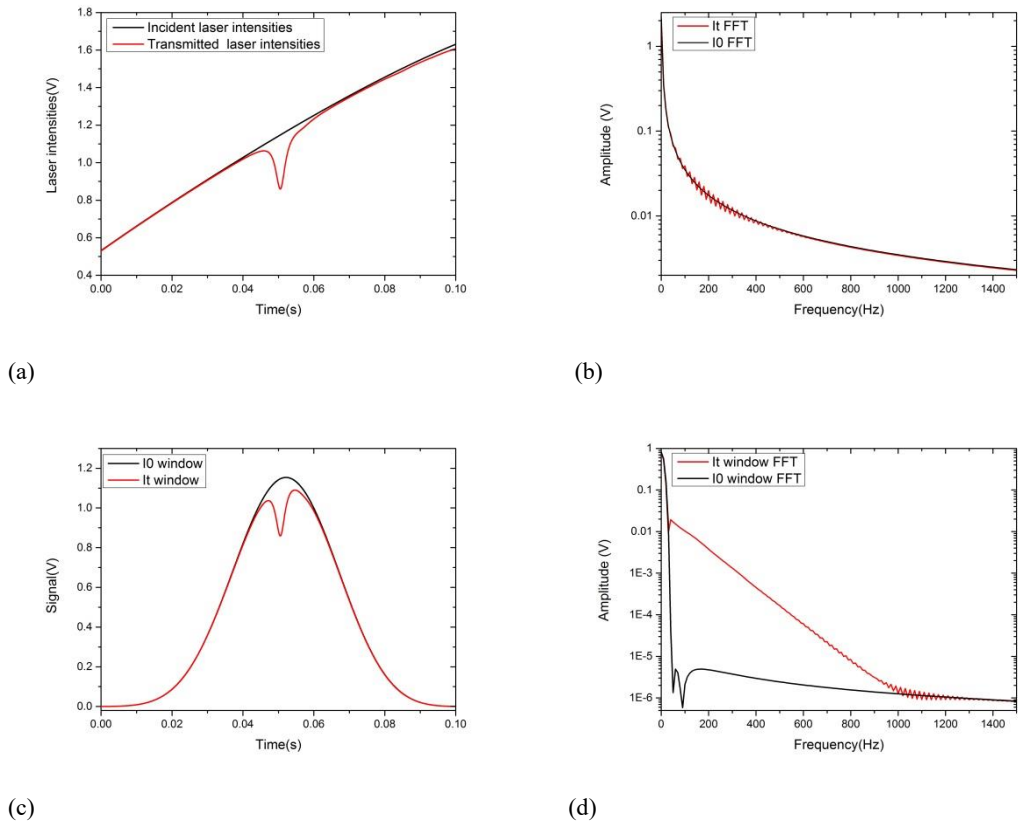

(a)　　　　　　　　　　　　　　　　　　(b)

(c)　　　　　　　　　　　　　　　　　　(d)

Fig. 5 (a) incident and transmitted laser intensity of a typical SDAS measurement, (b) FFT amplitude of incident and transmitted signal (c) windowed incident and transmitted signal (d) FFT of windowed incident and transmitted signal (simulation condition：2% $H_2O$；T=298 K；L=100 cm；P=1 atm；$f_0$(scanning frequency)=10Hz）

### 3.2 Theoretical analysis

Following assumptions are made for simplicity, (1) central absorption frequency is the same as center of scanning signal, (2) the laser intensity and wave number are first-order functions of time, (3) Lorenz line shape function is used, (4) only a single absorption line exists. Considering periodicity of the signal, it is only necessary to analyze the signal in one scanning cycle $-\frac{T}{2} \leq t \leq \frac{T}{2}$.

$$f_0 = \frac{1}{T} \tag{6}$$

$$v(t) = v_0 + \frac{2at}{T} \tag{7}$$

$$\emptyset[v(t)] = \frac{1}{\pi \Delta v_L} \frac{1}{1 + [(v(t) - v_0)/\Delta v_L]^2} \tag{8}$$

$$I_0(t) = \bar{I}_0 + \frac{2bt}{T} \tag{9}$$

$$\tau(t) = \exp(-SPXL\emptyset[v(t)]) \tag{10}$$

$$I_t(t) = I_0(t)\tau(t) = I_0(t)\exp(-SPXL\emptyset[v(t)]) \tag{11}$$

where T[s] is the scanning period, $f_0$[Hz] the scanning frequency, $v_0$[cm-1] the central absorption line frequency, v(t)[cm-1] the wave number at time t, $\Delta v_L$[cm-1] the full width half maximum of Lorenz line shape, $\tau(t)$ the transmittance. Figure 6 shows the incident and transmitted laser intensity within a period when the parameters take typical values.

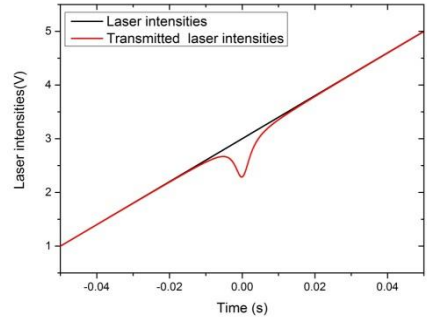

Fig. 6 Incident and transmitted laser intensity within a period for under a typical condition ($a = -1$; $b = 2$; $P = 1$atm; $X = 0.02$mol/mol; $L = 100$cm; $S = 0.0198$ cm$^{-2}$/atm; $T = 0.1$s; $\bar{I}_0 = 3$; $\Delta v_L = 0.04653575$cm$^{-1}$)

### 3.2.1 Fourier of SDAS signal

Incident laser intensity $I_0(t)$ is an odd function (when the constant term is ignored) with period of T, Fourier expansion of which is as below

$$I_0(t) = \bar{I}_0 + \sum_{n=1}^{\infty} (-1)^{n+1} \frac{2b}{n\pi} \sin(2\pi n f_0 t) \tag{12}$$

Transmittance $\tau(t)$ is an even function with period of T, Fourier expansion of which is,

$$\tau(t) = \sum_{k=0}^{\infty} H_k \cos(2\pi k f_0 t) \tag{13}$$

Where $H_k$ is the kth Fourier coefficient, defined as below,

$$H_k = \frac{1}{(1+\delta_{k0})\pi} \int_{-\pi}^{\pi} \tau\left(v_0 + \frac{a}{\pi}\theta\right) \cos(k\theta) \, d\theta \tag{14}$$

Multiply Equation (12) with (13) and rearrange the formula, the transmitted laser intensity can be obtained as below,

$$I_t(t) = H_0 \bar{I}_0 + \sum_{n=1}^{\infty} A_n \cos(2\pi n f_0 t) + \sum_{n=1}^{\infty} B_n \sin(2\pi n f_0 t) \tag{15}$$

where

$$A_n = H_n \bar{I}_0 \tag{16}$$

$$B_n = -\frac{bH_n}{2n\pi} + \sum_{\substack{k=0 \\ k \neq n}}^{\infty} (-1)^{n+k+1} \frac{bH_k}{\pi} \frac{2n}{n^2 - k^2} \tag{17}$$

### 3.2.2 Nuttall window

Nuttall window consists of cosine components, the expression in time domain is [15],

$$w(t) = \sum_{m=0}^{M-1} b_m \cos(2\pi m f_0 t) \tag{18}$$

Where, M is the number of terms, and $b_m$ have to satisfy constraint condition,

$$\sum_{m=0}^{M-1} b_m = 1 \tag{19}$$

$$\sum_{m=0}^{M-1} (-1)^m b_m = 0 \tag{20}$$

The coefficients of a typical Four-terms (i.e. M=4) Nuttall window used in this work are listed in Table 1.

Table 1 Nuttall window parameters

| Coefficients | 4 terms 3rd order | 4 terms 1st order | Minimum sidelobes of 3 terms | Minimum sidelobes of 4 terms |
|---|---|---|---|---|
| $b_0$ | 0.338946 | 0.355768 | 0.4243801 | 0.3635819 |
| $b_1$ | 0.481973 | 0.487396 | 0.4973406 | 0.4891775 |
| $b_2$ | 0.161054 | 0.144232 | 0.0782793 | 0.1365995 |
| $b_3$ | 0.018027 | 0.012604 |  | 0.0106411 |

### 3.2.3 Fourier analysis of windowed signal

Perform Fourier expansion on the Nuttall windowed incident laser intensity, we have

$$I_0(t)_{Window} = I_0(t) w(t) = \sum_{m=0}^{M-1} 2b_m \bar{I}_0 \cos(2\pi m f_0 t) \tag{21}$$

$$+ \sum_{m=0}^{M-1} b_m [\sum_{n=m+1}^{\infty} D_{n-m} \sin(2\pi n f_0 t) + \sum_{n=1-m}^{\infty} D_{n+m} \sin(2\pi n f_0 t)]$$

where,

$$D_n = (-1)^{n+1} \frac{2b}{n\pi} \quad (22)$$

Note when $n \geq M$, the high frequency components can be combined and the summation order can be switched.

$$I_0(t)_{Window} \atop n \geq M = \sum_{n=M}^{\infty} \sum_{m=0}^{M-1} b_m (D_{n-m} + D_{n+m}) \sin(2\pi n f_0 t) \quad (23)$$

Substitute Equation (22) into (23), we have,

$$I_0(t)_{Window} \atop n \geq M = \sum_{n=M}^{\infty} \sum_{m=0}^{M-1} \frac{(-1)^{m+n+1} b_m}{\pi} b \left( \frac{2}{n-m} + \frac{2}{n+m} \right) \sin(2\pi n f_0 t) \quad (24)$$

When $n \geq M$, amplitude of $n f_0$ harmonics of windowed incident laser intensity is

$$Amplitude_{I_0(t)Window}(n) = \sum_{m=0}^{M-1} \frac{(-1)^{m+n+1} b_m}{\pi} b \left( \frac{2}{n-m} + \frac{2}{n+m} \right) \quad (n \geq M) \quad (25)$$

amplitude of $n f_0$ harmonics of incident laser intensity without windowing operation is

$$Amplitude_{I_0(t)}(n) = (-1)^{n+1} \frac{2b}{n\pi} \quad (n \geq M) \quad (26)$$

Define the window function $R_0(n)$ of the incident laser intensity as below

$$R_0(n) = \frac{Amplitude_{I_0(t)Window}(n)}{Amplitude_{I_0(t)}(n)} = \sum_{m=0}^{M-1} (-1)^m b_m \left( \frac{n}{n-m} + \frac{n}{n+m} \right) \quad (n \geq M) \quad (27)$$

The norm of $|R_0(n)|$ is shown in Figure 7.

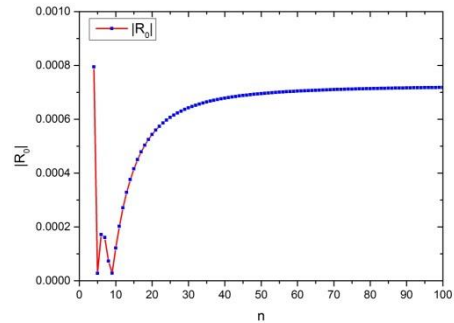

Fig. 7 Window function $R_0(n)$ of incident laser intensity

It can be seen from Figure 7, that amplitude of high frequency ($n \geq M$) harmonics of the incident laser intensity decreases more than 1000 times.

Perform Fourier expansion on the Nuttall windowed transmitted laser intensity, we have

$$I_t(t)_{Window} = I_t(t) w(t)$$
$$= \sum_{m=0}^{M-1} b_m \, H_0 \bar{I}_0 \cos(2\pi m f_0 t) + \sum_{m=0}^{M-1} \frac{b_m}{2} \sum_{n=m+1}^{\infty} A_{n-m} \cos(2\pi n f_0 t) \quad (28)$$

$$+ \sum_{m=0}^{M-1} \frac{b_m}{2} \sum_{n=1-m}^{\infty} A_{n+m} \cos(2\pi n f_0 t) + \sum_{m=0}^{M-1} \frac{b_m}{2} \sum_{n=m+1}^{\infty} B_{n-m} \sin(2\pi n f_0 t)$$

$$+ \sum_{m=0}^{M-1} \frac{b_m}{2} \sum_{n=1-m}^{\infty} B_{n+m} \sin(2\pi n f_0 t)$$

Similar to Equation (23), when $n \geq M$, the high frequency components of $I_t(t)_{Window}$ can be combined and the summation order can be switched,

$$\begin{aligned} I_t(t)_{Window} \atop n \geq M &= \sum_{n=M}^{\infty} \sum_{m=0}^{M-1} \frac{b_m}{2} (A_{n-m} + A_{n+m}) \cos(2\pi n f_0 t) \\ &+ \sum_{n=M}^{\infty} \sum_{m=0}^{M-1} \frac{b_m}{2} (B_{n-m} + B_{n+m}) \sin(2\pi n f_0 t) \end{aligned} \quad (29)$$

Define the term $\cos(2\pi n f_0 t)$ as X component of $n f_0$ harmonics, and term $\sin(2\pi n f_0 t)$ as Y component of $n f_0$ harmonics

The amplitude of X component of $n f_0$ harmonics of widowed transmitted laser intensity is,

$$\text{Amplitude}^X_{I_t(t)_{Window}}(n) = \sum_{m=0}^{M-1} \frac{b_m}{2}(A_{n-m} + A_{n+m}) \quad (n \geq M) \quad (30)$$

The amplitude of Y component of $n f_0$ harmonics of widowed transmitted laser intensity is,

$$\text{Amplitude}^Y_{I_t(t)_{Window}}(n) = \sum_{m=0}^{M-1} \frac{b_m}{2}(B_{n-m} + B_{n+m}) \quad (n \geq M) \quad (31)$$

The amplitude of X component of $n f_0$ harmonics of transmitted laser intensity without widow operation is,

$$\text{Amplitude}^X_{I_t(t)}(n) = A_n \qquad (n \geq M) \quad (32)$$

The amplitude of Y component of $n f_0$ harmonics of transmitted laser intensity without widow operation is,

$$\text{Amplitude}^Y_{I_t(t)}(n) = B_n \qquad (n \geq M) \quad (33)$$

Define the window function in X and Y directions of the transmitted laser intensity as $R^X(n)$ and $R^Y(n)$

$$R^X(n) = \frac{\text{Amplitude}^X_{I_t(t)_{Window}}(n)}{\text{Amplitude}^X_{I_t(t)}(n)} = \frac{\sum_{m=0}^{M-1} \frac{b_m}{2}(A_{n-m}+A_{n+m})}{A_n} \quad (n \geq M) \quad (34)$$

$$R^Y(n) = \frac{\text{Amplitude}^Y_{I_t(t)_{Window}}(n)}{\text{Amplitude}^Y_{I_t(t)}(n)} = \frac{\sum_{m=0}^{M-1} \frac{b_m}{2}(B_{n-m}+B_{n+m})}{B_n} \quad (n \geq M) \quad (35)$$

### 3.2.4 Significance of window function $R^X(n)$ and $R^Y(n)$

Transmitted laser intensity without widowing operation can be rewritten in the following form,

$$I_t(t) = I_0(t)\tau(t) = \left(\overline{I_0} + \frac{2bt}{T}\right)\tau(t) = \overline{I_0}\tau(t) + \frac{2bt}{T}\tau(t) \quad (36)$$

Apparently, the first term on the right hand side is an even function, while the second term is an odd function. Compare Equation (36) and (15), we have

$$\bar{I}_0\tau(t) = H_0\bar{I}_0 + \sum_{n=1}^{\infty} A_n \cos(2\pi nf_0 t) \qquad (37)$$

$$\frac{2bt}{T}\tau(t) = \sum_{n=1}^{\infty} B_n \sin(2\pi nf_0 t) \qquad (38)$$

Windowed transmitted laser intensity signal can be rewritten as Equation (39) below,

$$I(t)_{Window} = I_0(t)\tau(t)w(t) = \bar{I}_0\tau(t)w(t) + \frac{2bt}{T}\tau(t)w(t) \qquad (39)$$

Similarly

$$\bar{I}_0\tau(t)w(t) \bigg|_{n \geq M} = \sum_{n=M}^{\infty} \sum_{m=0}^{M-1} \frac{b_m}{2}(A_{n-m} + A_{n+m})\cos(2\pi nf_0 t) \qquad (40)$$

$$\frac{2bt}{T}\tau(t)w(t) \bigg|_{n \geq M} = \sum_{n=M}^{\infty} \sum_{m=0}^{M-1} \frac{b_m}{2}(B_{n-m} + B_{n+m})\sin(2\pi nf_0 t) \qquad (41)$$

According to Equation (38-41), $A_n$ mainly contains the information of $\bar{I}_0\tau(t)$, and $B_n$ mainly contains information of $\frac{2bt}{T}\tau(t)$. Amplitude of X component of $nf_0$ harmonics of widowed transmitted laser intensity depends on $A_n$, and amplitude of Y component depends on $B_n$. The amplitude of X component reflects the contribution of the average amplitude of the original intensity signal (baseline) to the transmitted light intensity, and the amplitude of X component reflects contribution of the trend (slope) of the original light intensity signal (baseline) to the transmitted light intensity.

Thus, $R^X(n)$ represents the inhibition effect of the windowing operation to the average amplitude of the incident intensity signal (baseline), and $R^Y(n)$ represents the inhibitory effect of windowing on the trend (slope) of the original intensity signal (baseline). Variations of $R^X(n)$ and $R^Y(n)$ are show in Figure 8.

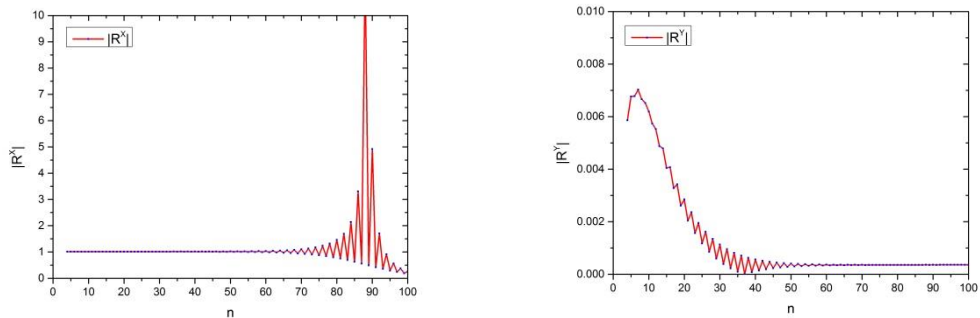

Fig. 8 Window function $R^X(n)$ and $R^Y(n)$

According to Figure 8, the high frequency components of the amplitude of the X-direction harmonics of the absorption hardly change between $4f_0$ to $70f_0$, while those of the Y-direction decreases to be less than 0.7% of original value. Especially the components larger than $30f_0$, the amplitude decreases to less than 0.1%. Therefore, the amplitude of Y direction harmonics attenuates substantially in high frequency section ($n \geq M$), while the amplitude of X direction does not change much. This indicates that the windowing of the transmitted light intensity signal preserves the absorption information $A_n$, and meanwhile effectively suppresses the error due to the inaccuracy of fitted baseline slop. The deviation from the mean amplitude $\bar{I}_0$ in the baseline can be removed by normalization. Extract the constant terms in Equation (28),

$$C = \bar{I}_0 \left( b_0 H_0 + \frac{b_1}{2} H_1 + \frac{b_2}{2} H_2 + \frac{b_3}{2} H_3 \right) \quad (42)$$

Normalize Equation (30 with Equation (42), we have the normalized amplitude of the nf0 harmonics in X direction of the windowed absorption signal, which is independent with $\bar{I}_0$.

$$\text{Amplitude}_{I_t(t)\text{Window}}^{X\ \text{normalized}}(n) = \frac{\sum_{m=0}^{M-1} \frac{b_m}{2}(H_{n-m} + H_{n+m})}{\sum_{m=0}^{M-1} \frac{b_m}{2-\delta_{m0}} H_m} \quad (n \geq M) \quad (43)$$

Thus, we obtain the Equation (43) which is independent of the slope and amplitude of the baseline and contains the absorption information. Considering the component of the high frequency portion of the nf0 harmonic Y direction is also suppressed due to windowing, Equation (43) can be obtained through digital filtering of windowed transmitted light intensity signal. Then the temperature and concentration information of gas can be also obtained. In the practical application of the BF-SDAS method, the baseline can be obtained by conventional SDAS fitting, to minimize the slope error of the baseline and further reduces the measurement error.

### 3.2.5 Determination of the absorption center

Since the high frequency components of nf0 harmonics in Y direction is suppressed, an approximation of Equation (27) can be written as,

$$\begin{aligned} I_t(t)_{\text{Window}} &\cong \sum_{n=M}^{\infty} \sum_{m=0}^{M-1} \frac{b_m}{2}(A_{n-m} + A_{n+m})\cos(2\pi n f_0 t) \\ n \geq M & \\ &= \sum_{n=M}^{\infty} \sum_{m=0}^{M-1} \frac{b_m \bar{I}_0}{2}(H_{n-m} + H_{n+m})\cos(2\pi n f_0 t) \quad (n \geq M) \end{aligned} \quad (44)$$

When t=0, $I_t(t)_{\text{Window}}$, $n \geq M$ gets the maximum, i.e. the sum of the harmonics at the absorption center has the maximum value. Therefore, the maximum value of the sum of the harmonics can be used to calibrate the position of the absorption center.

## 4 Experimental validation

### 4.1 Experimental setup

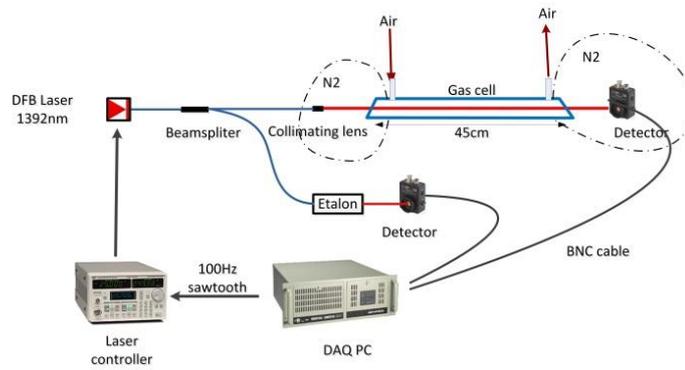

Fig. 9 Experimental setup of SDAS

Figure 9 illustration of experimental setup. First, set wavelength of the laser output (NEL NLK1E5GAAA) to the absorption line wavelength. A function generator is used to generate the sawtooth signal to modulate input current of the laser generator. The sawtooth wave has a frequency of 100 Hz and an amplitude of 2 Vpp. The wavelength of laser generator is to scan the selected absorption region lineshape. The output laser is then split into two parts: one transmits through the gas object until radiated on the detector (PDA-20CS-EC), and the other one passes through an Etalon and arrives at another detector (PDA-20CS-EC). The etalon can convert the time-domain intensity signal to the frequency-domain intensity signal, the simplest form of which is the Fabry-Perot interferometer. The two detected signals are denoted as $I_t$ and $I_v$ respectively. Comparing these two signals, the absorption on the frequency domain can be obtained. Further spectral analysis would yield concentration information of the target species.

This work uses 7185.6cm$^{-1}$ absorption line of $H_2O$ molecular, which is widely used in laser diagnostics, to detect the $H_2O$ molar concentration. Detailed information of spectral parameters, spectral line position, line intensity and lower level energy can be found in Table 2, same as Sun [3].

Table 2 spectral parameters, spectral line position, line intensity and lower level energy

| Line index | Wave number $v_0$ [cm-1] | S at 296K [atm-1cm-2] | E" [cm-1] |
|---|---|---|---|
| 1 | 7185.60 | 0.0195 | 1045 |
| 2 | 7185.39 | 0.00121 | 447 |

**4.2 Results and discussion:**

**4.2.1 Results**

Figure 10 shows the fitting results of the traditional SDAS method and the absorbance. Due to the fitting error (~4%) of the baseline, the final concentration calculated with SDAS method is $X_{SDAS}$=8.175×10$^{-3}$ mol/mol. As a comparison, Figure 11 shows the results of BF-SDAS method. Figure 11(a) shows the signal obtained by windowing the same data as in Figure 10(a). The windowed transmission intensity signal corresponding to Equation (40) is obtained by digital low pass and digital bandpass filtering, as shown in Figure 11(b). The fitting error is within 0.4%, which is much better than the SDAS results. The final concentration calculated with BF-SDAS is then $X_{SDAS}=8.426\times10^{-3}$ mol/mol. Besides the absorption measurement, a hygrothermograph (HMT333) was used to measure the humidity and temperature of experimental environment. When working at 20 °C and humidity of 0%~90%, the humidity and temperature precision are 1% and 0.2 °C respectively. The measured temperature is 22.1 °C and humidity is 33%. Considering the accuracy of the hygrothermograph, apparently the BF-SDAS method is more convincing. The measurement results of $H_2O$ concentration are listed in Table 3.

Table 3 $H_2O$ concentration results

| Method | Concentration (mol/mol) | Standard deviation σ/(mol/mol) | error (mol/mol) |
|---|---|---|---|
| HMT333 | 0.008665 | | ±0.0003720 |
| SDAS | 0.008175 | 0.0001721 | |
| BF-SDAS | 0.008426 | 0.0001532 | |

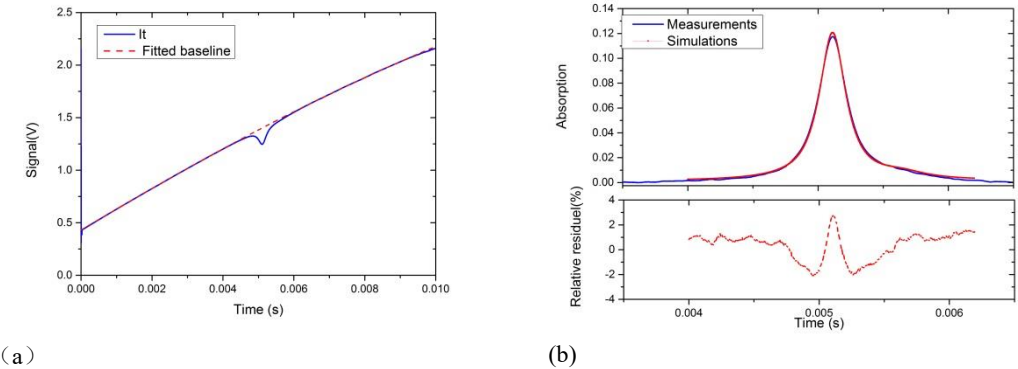

(a)　　　　　　　　　　　　　　　　　(b)

Fig. 10 Measurement results of traditional SDAS

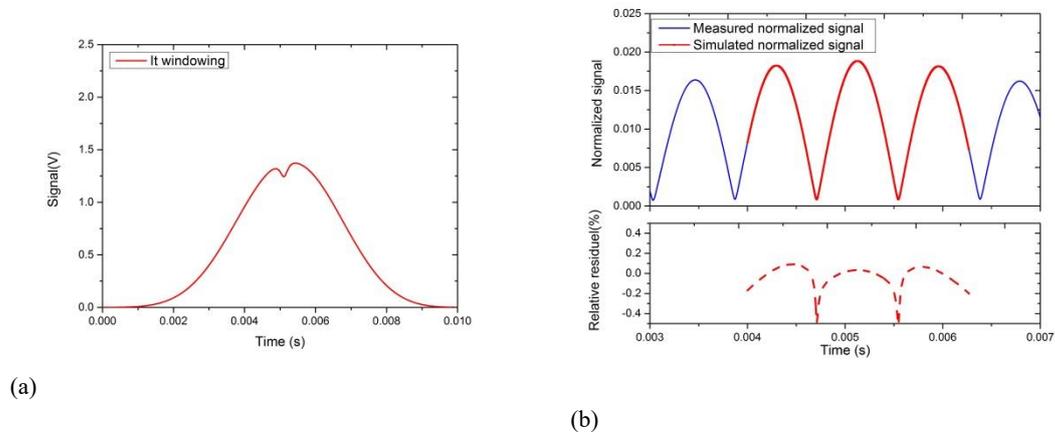

(a)

(b)

Fig. 11 Measurement results of BF-SDAS

### 4.2.2 Analysis of two methods

Because of the inevitable fitting errors in the baseline fitting process, mainly shift of the baseline and variation of the baseline slope, the effect of the baseline fitting error on the results is discussed below.

During the SDAS-based baseline fitting process, different offset was added into the fitted baseline. The magnitudes of offset in terms of percentage of average amplitude are 0, ± 0.1%, ± 0.2%, ± 0.5%, respectively. The corresponding results of SDAS method and BF-SDAS method are shown in Fig. 12 (a) and Fig. 12 (b), respectively. The absolute and relative change of $H_2O$ concentration calculated by the two methods are shown in Fig. 12(c) and (d). Obviously, the fitting result of the SDAS method has a significant dependent on the baseline offset amount, while the BF-SDAS method is almost independent on the baseline offset. The absolute and relative change of the $H_2O$ concentration obtained by the two methods have the same trend, but SDAS calculated the concentration with ±30% relative change, while the BF-SDAS method to calculate the concentration with only ± 0.01% of the relative change.

In the SDAS-based baseline fitting process, the slope of the fitted baselines was varied by different amount, 0, ±0.1%, ±0.2%, and ±0.5 %, in terms of the percentage of slope change with respect to the original fitted slope. The results are shown in Figure 13(a), and the relative change of $H_2O$ concentration calculated by the two methods is shown in Fig. 13 (b). It is seen that the fitting result of SDAS method is significantly related to the slope change of baseline, while BF-SDAS method has no dependence on the slope change. The relative change of $H_2O$ concentration calculated by the two methods has the same trend. The relative concentration variation of $H_2O$ calculated by SDAS method varies within ±15%, while the concentration calculated by BF-SDAS method is just ±0.005%.

The above analysis shows that the BF-SDAS method is not sensitive to the baseline fitting error compared with the traditional SDAS method. Therefore it substantially improves the stability and precision of the measurement results. It is worth noting that the BF-SDAS method still needs an initial baseline estimate for simulation in the actual calculation process, though the baseline accuracy requirement is greatly reduced.

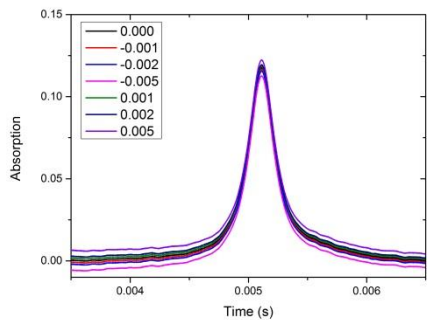

(a)

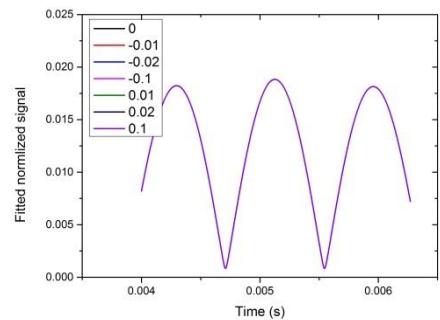

(b)

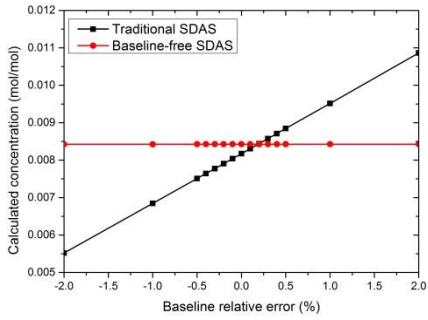 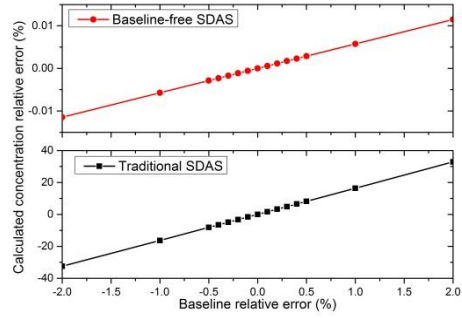

(c)                                       (d)

Fig. 12 Effect of baseline shift

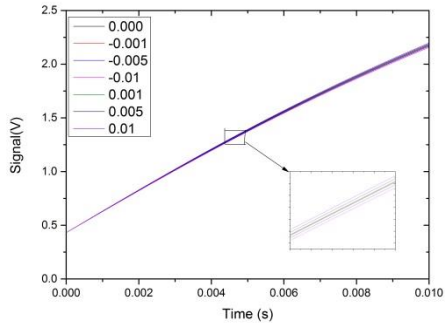 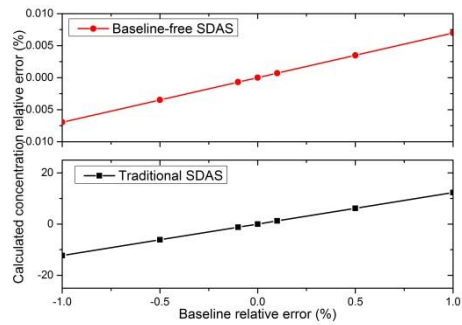

(a)                                       (b)

Fig. 13 Effect of baseline slope

## 5. Conclusion

In this paper, a baseline-free direct-absorption-spectroscopy (BF-SDAS) method is proposed to eliminate the dependence of scanned-wavelength direct absorption spectroscopy (SDAS) on pressure and temperature parameters in combustion diagnostics. First, the BF-SDAS method is theoretically analyzed based on the Fourier method, and the accuracy of the method is validated by measuring the concentration of $H_2O$ in a controlled and static gas absorption cell at room temperature.

Compared with the traditional direct absorption method, the BF-SDAS method has the following advantages: (1) It overcomes the sensitivity of the traditional direct absorption method to the baseline fitting error, (2) BF-SDAS is more suitable for high pressure combustion environment, while the pressure broadening in high pressure environment introduces spectral line overlapping and makes the baseline fitting difficult or even impossible, (3) The BF-SDAS method can obtain correct result when scanned spectrum is not complete, which is useful for cases when the gas state under inspection varies intensively, while the traditional SDAS cannot be applicable.

In summary, the traditional SDAS only applies to occasions of appropriate absorbance, relatively low pressure, small modulation frequency, and isolated absorption line, while BF-SDAS can be well adapted to high pressure and high modulation frequency cases, because of its insensitivity to baseline fitting error.